\documentclass[fleqn,twoside]{article}
\usepackage{espcrc2}

\usepackage{graphicx}
\newcommand{\AmS}{{\protect\the\textfont2
  A\kern-.1667em\lower.5ex\hbox{M}\kern-.125emS}}

\newcommand{\refeq}[1]{Eq.~(\ref{eq:#1})}
\newcommand{\reffig}[1]{Fig.~\ref{fig:#1}}
\newcommand{\refsec}[1]{Sec.~\ref{sec:#1}}

\newcommand{\Deg}{^{\circ}}
\newcommand{\Sem}{S_{\rm em}}

\newcommand{\Smu}{S_{\mu}}
\newcommand{\Xmax}{X_{\rm max}}
\newcommand{\Xmaxavg}{\langle X_{\rm max} \rangle}

\newcommand{\Nmu}{N_{\mu}}
\newcommand{\Nmufit}{N_{\mu\:\rm fit}}

\newcommand{\psize}{72mm}

\title{Measurements of the Muon Content of UHECR Air Showers with 
the Pierre Auger Observatory}

\author{Fabian Schmidt\address{Department of Astronomy \& Astrophysics,
University of Chicago, Chicago, IL 60637, USA}\thanks{fabians@oddjob.uchicago.edu},
for the Pierre Auger Collaboration\address{Observatorio Pierre Auger, 
Av. San Mart\'in Norte 304, (5613) Malarg\"ue, Mendoza, Argentina}}
       
\begin{document}

\begin{abstract}
The Pierre Auger Observatory, recently completed, has been operational 
since 2004. As a hybrid experiment, it allows for a wide range of 
measurements of UHECR-induced extensive air showers (EAS), including 
measurements of the EAS particle content on ground which is sensitive 
to high-energy hadronic interactions. We present the results of several independent 
measurements of the EAS muon content on ground in Auger data at a primary 
energy of 10~EeV. We discuss implications on high-energy 
hadronic interaction models and cosmic ray composition. 
\end{abstract}

% typeset front matter (including abstract)
\maketitle

%%%%%%%%%%%%%%%%%%%%%%%%%%%%%%%%%%%%%%%%%%%%%%%%%%%%%%%%%%%%%%%%%%%%%%%%%%%
\section{Introduction}
\label{sec:intro}

Cosmic ray interactions at ultra-high energies offer unique insights
into particle collisions at center-of-mass energies exceeding 100~TeV.
Due to their very small flux at these energies, cosmic rays can only
be detected indirectly via the extensive air showers (EAS) they induce
in the atmosphere. Hence, in order to infer the mass and energy
of the primary cosmic rays as well as to gain insight into the physics of 
the high-energy collisions, one has to rely on detailed simulations
of air showers. For this purpose, hadronic multiparticle production 
has to be simulated at
energies exceeding by far those accessible at man-made accelerators
and in phase space regions not covered in collider
experiments. The secondary muons produced in these collisions which can be 
detected directly on ground, are a tracer of high-energy hadronic interactions.
For this reason, the number of muons produced in simulated air showers depends 
strongly on the adopted hadronic
interaction models \cite{Knapp:2002vs}. In addition, muons are a sensitive
indicator of the primary composition, as, for example, iron showers produce
about 40\% more muons than protons.

In this work we will employ {\it air shower universality} in order to
determine the number of muons in air showers detected with Auger, using
several independent methods. Air shower universality 
\cite{BilloirCORSIKASchool05,univpaper} states
that the average properties of air showers can be completely characterized 
by the primary energy, stage of shower evolution, and overall muon 
normalization, when viewing the shower around or after shower maximum.
In particular, the electromagnetic particle content of the
shower is already determined by its energy and stage of evolution.
Physically, this is due to the fact that the electromagnetic
cascade involves dozens of particle generations and of order $10^{9}-10^{10}$
particles, which washes out the details of high-energy hadronic interactions.

In order to infer the muon density of showers at a given energy,
one has to rely on an energy scale of the surface detector. Using
the universality of the electromagnetic contribution to the signal
measured by the Auger surface detectors, $S(1000)$, we are able to determine
an energy scale independent of the fluorescence energy measurement.
The muon density at ground which we infer using the surface detector and,
independently, using hybrid events, does not rely on
assumptions on the primary cosmic ray composition. This allows for a
direct test of the predictions of hadronic interaction models.

%%%%%%%%%%%%%%%%%%%%%%%%%%%%%%%%%%%%%%%%%%%%%%%%%%%%%%%%%%%%%%%%%%%%%%%%%%%
\section{Parameterization of the average surface detector signal using universality}
\label{sec:univ}

Universality features of the longitudinal profile of showers have been
studied by several authors. Here we exploit
shower universality features to predict the surface detector signal
expected for Auger Cherenkov tanks from the electromagnetic and
muonic shower components at 1000\,m from the shower core. In the
following, we give a brief outline of the parameterization. 
For a detailed description, see \cite{univpaper}.

A library of proton and iron showers covering the energy range from
$10^{17}$ to $10^{20}$\,eV and zenith angles between $0^\circ$ and
$70^\circ$ was generated with CORSIKA 6.5 \cite{Heck98a} and the
hadronic interaction models QGSJET II.03 \cite{Ostapchenko:2005nj} and
FLUKA \cite{Fasso01a}. For comparison, a smaller set of showers was
simulated with the combinations QGSJET II.03/GHEISHA
\cite{Fesefeldt85a} and SIBYLL 2.1/FLUKA
\cite{Fletcher:1994bd,Engel99a}. Seasonal models of the Malarg\"ue
molecular atmosphere were used \cite{Keilhauer:2004jr}. The detector 
response is calculated using look-up tables derived from a detailed 
GEANT4 simulation of the Auger surface detectors, which has been
shown to match the data well \cite{Auger-ICRC07-Henry-Yvon}. Note that 
since the Auger detectors are calibrated against the cosmic ray muon
background, we only rely on the correct simulation of the 
signal generated by electromagnetic particles relative to that of the muons.

\begin{figure}[ht!]
\includegraphics[width=\psize]{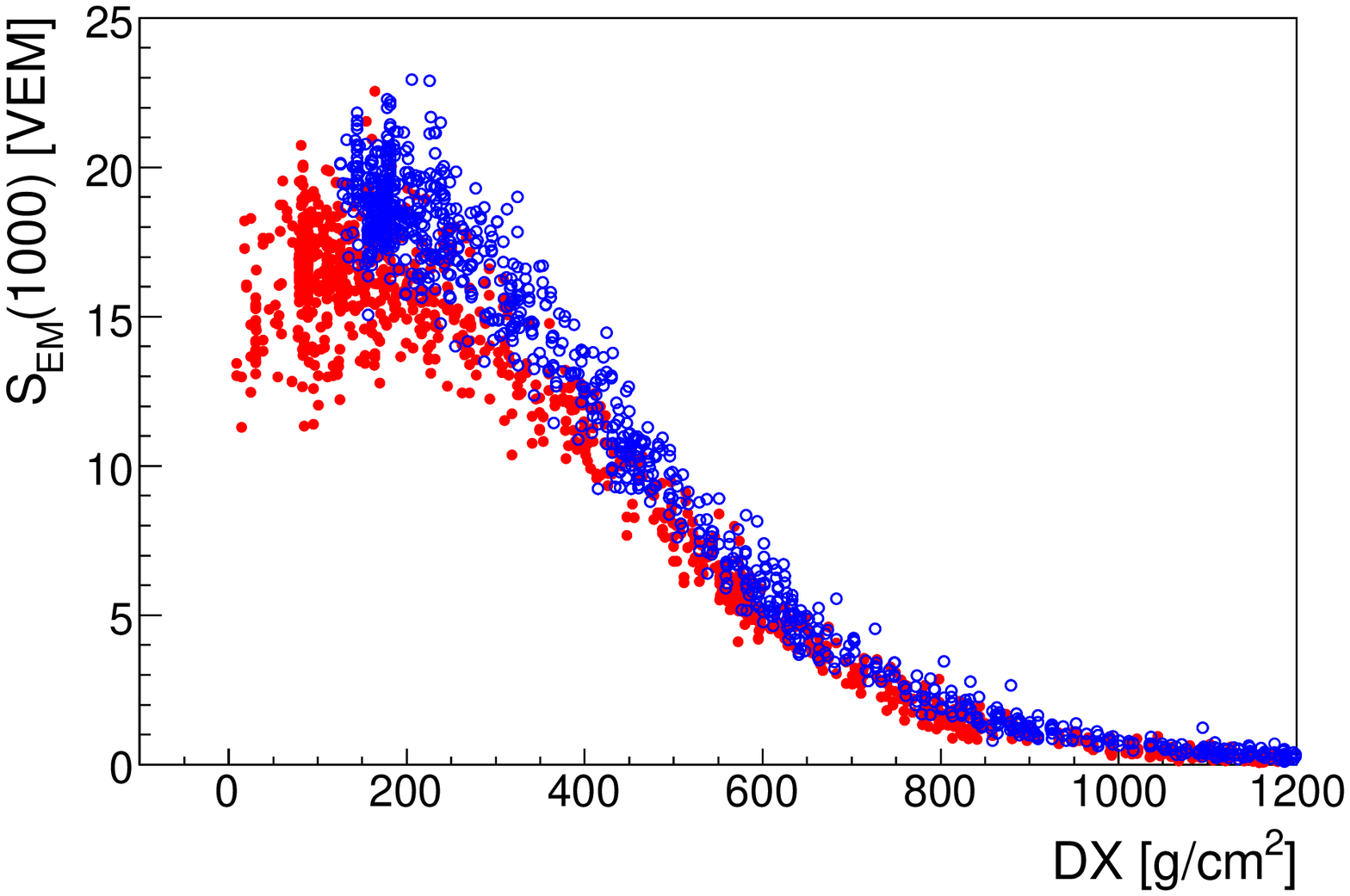}\\
\includegraphics[width=\psize]{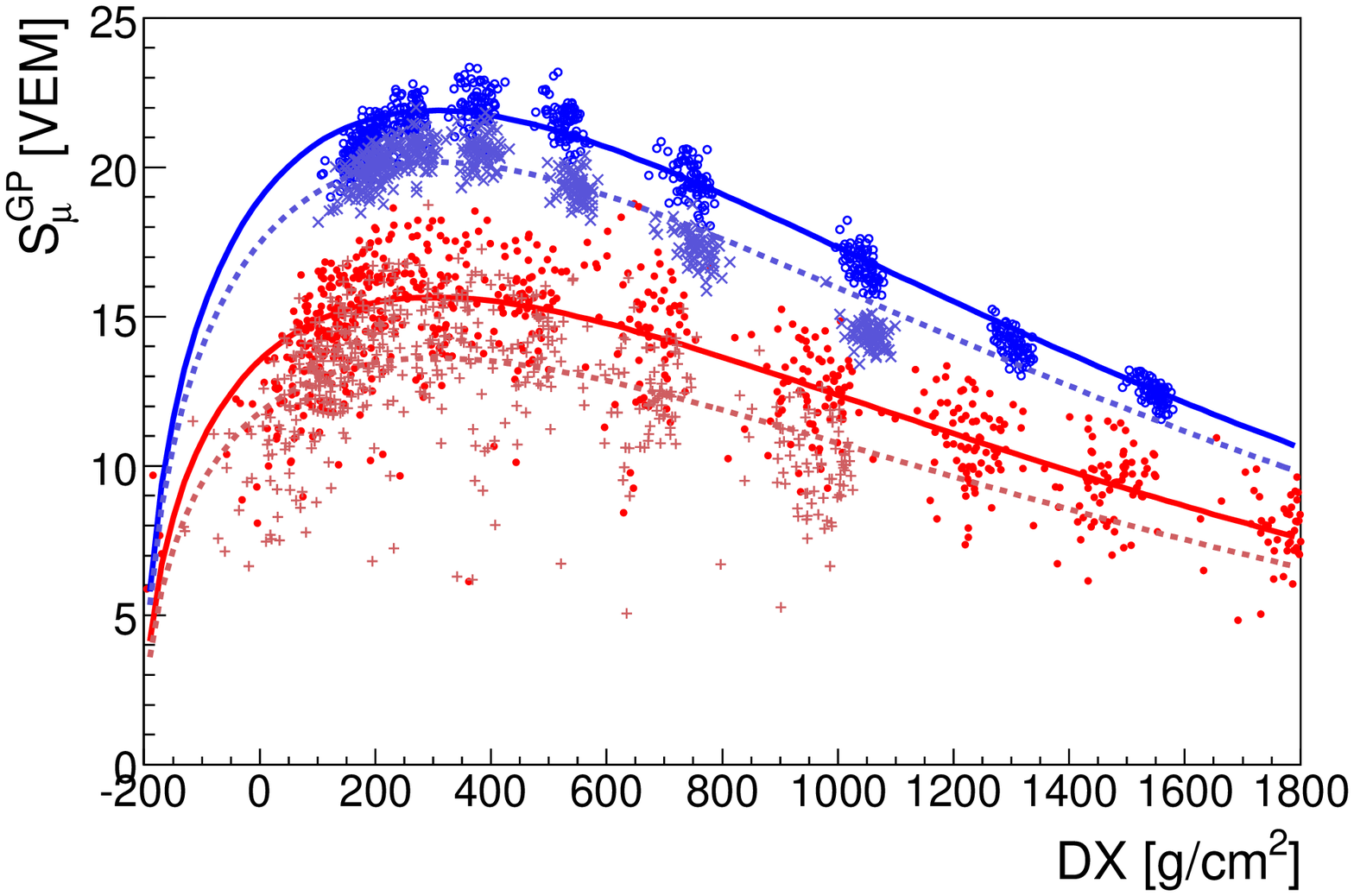}
%\vspace*{-0.2cm}
\caption{\small{{\it Upper panel:} Simulated electromagnetic shower plane signals 
(excluding muon decay products) at 1000~m for proton (red dots) and 
iron showers (blue circles) at $10^{19}$~eV ($\theta=0-60\Deg$; using QGSJetII
and Fluka) as a function of $DX$.
{\it Lower panel:} Simulated muon signals at 1000~m (including
muon decay products) vs. $DX$ for the same simulated showers, and proton and
iron showers simulated using Sibyll (crosses). The lines
show joint Gaisser-Hillas type parameterizations of all primaries and models.
\label{fig:Sparam}}}
\vspace*{-0.5cm}
\end{figure}

Air shower universality states that the electromagnetic component of the 
predicted surface detector signal
$S(1000)$ at the lateral distance of 1000\,m depends only on primary
energy and the stage of shower evolution, and not on the primary particle
and hadronic model assumed. We define the electromagnetic component of the 
signal  as that of electrons, positrons, and gamma rays excluding muon decay 
products, and measure shower evolution in terms of the
distance between the shower maximum and the ground 
$DX= X_{\rm ground} - X_{\rm max}$, measured along the shower axis.
In fact, universality is slightly violated, and the electromagnetic contribution
to $S(1000)$ of proton and iron showers differ by about 12\% \cite{univpaper}, 
see \reffig{Sparam}. 
The dependence on the hadronic model is much smaller, about 5\%.

%%%%%%%%%%%%%%%%%%%%%%%%%%%%%%
%%%%%%%%%%%%%%%%%%%%%%%%%%%%%%

The muon contribution to $S(1000)$, which includes muon decay products
in our definition, depends strongly on the primary particle
(40\% difference between proton and iron) and the hadronic model 
(\reffig{Sparam}). However, the functional dependence on $DX$ is universal.

After accounting for geometrical effects such as the projected tank
surface area, the electromagnetic shower signal (averaged over proton and iron)
is parameterized as function of the energy $E$, distance to shower
maximum $DX$, and zenith angle $\theta$. The difference between proton
and iron shower profiles is included in the calculation of the
systematic uncertainties later on. Similarly, the universal shape of the muon
signal profile is parameterized simultaneously using a Gaisser-Hillas function
for all models and primaries (\reffig{Sparam}),
leaving only normalization factors free. The expected detector
signal at 1000\,m can then be written as
\begin{eqnarray}
S_{\rm param}(E,\theta, X_{\rm max},\Nmu) = S_{\rm em}(E,\theta,DX(\Xmax)) & &
\nonumber\\
+ \Nmu\: S_\mu^{\rm QGSII,p} (10\,{\rm EeV}, \theta, DX(\Xmax)), & &
\label{eq:S1000}
\end{eqnarray}
where $S_\mu^{\rm QGSII,p}$ is the parametrized muon signal adopting the
normalization for proton-QGSJET II at 10\,EeV, so that
$\Nmu$ is the number of muons relative to 
that of QGSJET-II proton showers at 10\,EeV.

%%%%%%%%%%%%%%%%%%%%%%%%%%%%%%%%%%%%%%%%%%%%%%%%%%%%%%%%%%%%%%%%%%%%%%%%%%%
\section{Measuring $\Nmu$ from the surface detector}
\label{sec:CIC}

Within the current statistics, the arrival directions of
high-energy cosmic rays around 10~EeV are found to be isotropic, allowing us to apply
the {\it constant intensity} method to determine the muon signal
contribution. Dividing the surface detector data into equal exposure
bins, the number of showers with $S(1000)$ greater than a given
threshold $S_{\rm param}(10\:\mbox{EeV},\theta,\Xmaxavg,\Nmu)$, should be the same for each bin.
\begin{equation}
\left. \frac{dN_{\rm ev}}{d\sin^2 \theta}\right|_{S(1000) > S_{\rm param}(E,\theta,\langle X_{\rm max} \rangle, 
\Nmu)} = \rm const.
\label{eq:CIC-muons}
\end{equation} 
Using the independently measured mean depth of shower maximum $\langle
X_{\rm max} \rangle$ \cite{Auger-ICRC07-Unger}, the only remaining free
parameter in Eq.~(\ref{eq:CIC-muons}) is the relative number of muons
$\Nmu$. For a given energy $E$, $\Nmu$ is
adjusted to obtain a flat distribution of events in $\sin^2\theta$.

More precisely, scanning through a range of $\Nmu$ values, we calculate the $\chi^2$/dof 
of the event histogram relative to a flat distribution in $\sin^2\theta$. 
We then fit a two-branch parabola around the minimum of $\chi^2(\Nmu)$,
which results in a best-fit $\Nmufit$ and its asymmetric error, 
$\sigma^\pm_{\Nmu}$.

\begin{figure}[ht!]
%\vspace{9pt}
\includegraphics[width=\psize]{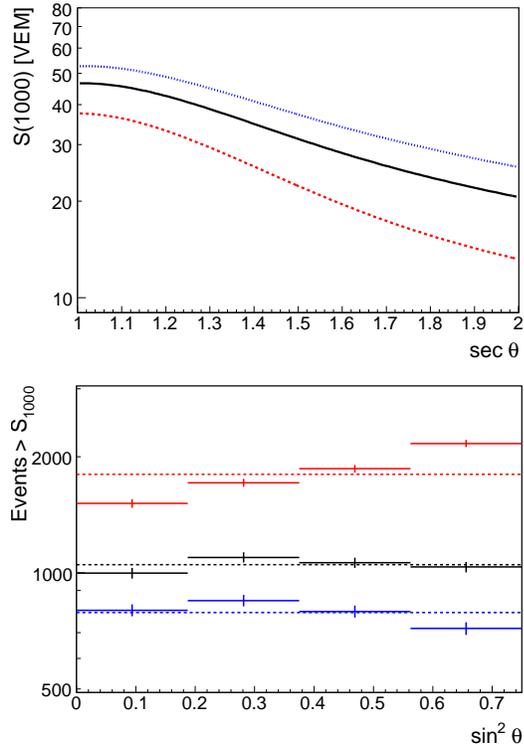}
\caption{\small{{\it Upper panel:} the signal parameterization \refeq{S1000} vs. 
$\sec\:\theta$ for different $\Nmu$ (black/solid$-$1.6, red/dashed$-$1.0, 
blue/dotted$-$2.0). {\it Lower panel:} 
histograms of number of Auger events above the parameterized signal in equal
exposure bins, obtained for the same $\Nmu$ as shown in the upper panel.
The black histogram is for $\Nmu=1.6$, the best-fit value found in the data
(see text).
\label{fig:cic-results}}}
\vspace*{-0.5cm}
\end{figure}

The sensitivity of this method to the muon number parameter in
Eq.~(\ref{eq:S1000}) is illustrated in Fig.~\ref{fig:cic-results}. Clearly,
$\Nmu=1$ (top histogram in the lower panel of \reffig{cic-results}) is ruled
out by the data. The
best description of the data above 10\,EeV requires $\Nmufit = 1.60$. However, 
we have to take into account shower-to-shower
fluctuations and the finite resolution of $S(1000)$ which lead to a slightly 
biased $\Nmu$ measurement in this method.

%%%%%%%%%%%%%%%%%%%%%%%%%%%%%%
%%%%%%%%%%%%%%%%%%%%%%%%%%%%%%

In order to estimate this bias, we have simulated
a large number of Monte Carlo realizations of SD data sets of the same size as the
Auger data set, distributed according to the observed cosmic ray spectrum and 
assuming different primary compositions (pure proton, iron, or mixed composition),
and different ``true'' $\Nmu$. For each shower, $\Xmax$ is obtained from 
distributions which closely
match the observed $\Xmax$ distributions in the fluorescence data.
The electromagnetic and muon signals are fluctuated according to the 
model predictions \cite{univpaper}. Note that the magnitude of fluctuations in
$\Xmax$ and $\Nmu$ are robust predictions which only depend on the 
primary particle.
\refeq{S1000} is then used to calculate the signal at 1000 m from the shower 
core, $S(1000)$, which is smeared with an experimental reconstruction 
accuracy \cite{Ave-ICRC}. A realistic trigger probability is also applied.

\begin{figure}[t]
\includegraphics[width=\psize]{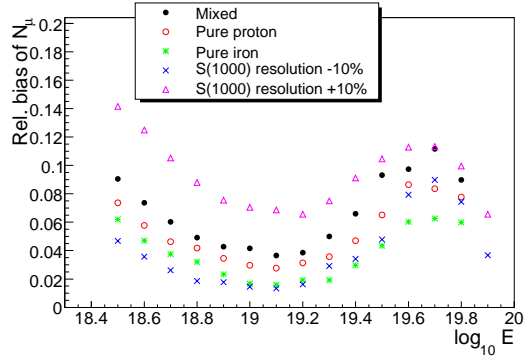}
\caption{\small{Relative bias on $\Nmu$ measured with the constant intensity 
method, determined from Monte Carlo simulations of Auger data sets.
The different points show various assumptions on the primary composition
and the experimental resolution of the $S(1000)$ reconstruction.
\label{fig:bias}}}
\vspace*{-0.5cm}
\end{figure}

For each simulated data set, we then apply exactly the same analysis
as used on the real data. We found that $\Nmu$ is systematically 
{\it overestimated} by 3$-$6\%. This bias is due to a combined action
of shower-to-shower fluctuations and detector resolution which depends
on the signal size. We have found that the bias in $\Nmu$ is independent
of the ``true'' value of $\Nmu$ adopted, and primarily depends on
the size of reconstruction uncertainties on $S(1000)$
and the primary composition. \reffig{bias} shows the relative bias on $\Nmu$
as a function of primary energy, for different assumed compositions and
increased/decreased detector resolution. Clearly, the bias is quite robust 
to rather extreme
changes in those assumptions.

We then subtract the bias expected for a mixed composition (black dots
in \reffig{bias}) from the measured 
$\Nmufit$, obtaining a corrected $\Nmu$ of:
%\begin{eqnarray}
%\Nmu(10\:\mbox{EeV}) &=& 1.53 +0.09/ -0.07 (\rm stat.)\nonumber \\
%& &  +0.21/-0.11 (\rm syst.).
%\label{eq:NmuCIC}
%\end{eqnarray}
\begin{equation}
\Nmu(10\:\mbox{EeV}) = 1.53^{+0.09}_{-0.07}\,({\rm stat.})
{}^{+0.21}_{-0.11}\,({\rm syst.})
\label{eq:NmuCIC}
\end{equation}
Three main sources of systematics enter in the determination of $\Nmu$ using 
the constant intensity method:
the uncertainty on the electromagnetic signal due to universality violation
(\refsec{univ}); the uncertainty on $\Xmaxavg$ at this energy; and the
uncertainty on the bias of $\Nmu$ determined from the Monte Carlo simulations.

In order to quantify the first two systematics, we measure $\Nmu$ using an
electromagnetic signal rescaled by $\pm$6\%, bracketing the observed
universality violation, 
and measuring $\Nmu$ with varying $\Xmaxavg$, adopting the systematic and
statistical uncertainty reported by Auger \cite{Auger-ICRC07-Unger}. 
For the uncertainty
on the bias, we adopt $\pm 3$\%. Summing the different contributions in 
quadrature, we obtain the total systematic error on $\Nmu$ quoted in
\refeq{NmuCIC}.

Once $\Nmu$ at 10~EeV is measured, and using the measured mean depth of 
shower maximum, the signal size at $\theta=38^\circ$ can be calculated:
%\begin{eqnarray}
%S_{38}(10\: {\rm EeV}) = 38.9 +1.4/-1.2 ({\rm stat.}) \nonumber\\
%+1.6/-1.8 \, ({\rm sys.}) \, {\rm VEM}.
%\label{eq:energy-scale}
%\end{eqnarray}
\begin{equation}
S_{38}(10\: {\rm EeV}) = 38.9^{+1.4}_{-1.2}\,({\rm stat.})
{}^{+1.6}_{-1.8}\,({\rm sys.})\,\rm VEM\:
\label{eq:energy-scale}
\end{equation}
Within the uncertainties, this value of $S_{38}$ is consistent
with the energy scale from fluorescence detector measurements, whose 
systematic uncertainty (22\% \cite{PRL}) is
dominated by the uncertainty on the fluorescence yield.
\refeq{energy-scale} corresponds to assigning showers a $\sim
26$\% higher energy than done in the fluorescence detector-based Auger
shower reconstruction ($E = 1.26\: E_{\rm FD}$).

%%%%%%%%%%%%%%%%%%%%%%%%%%%%%%%%%%%%%%%%%%%%%%%%%%%%%%%%%%%%%%%%%%%%%%%%%%%
\section{Measuring $\Nmu$ from hybrid events}
\label{sec:hybrid}

The Pierre Auger Observatory has the unique capability of measuring hybrid events
which have been simultaneously detected by the surface detector array
and fluorescence telescopes. For each of these events, a calorimetric
measurement of the energy $E$ and depth of shower maximum $\Xmax$ are 
available, in addition to the signal $S(1000)$ from the surface array.
Hence, we can use our parameterization of the universal electromagnetic
signal $\Sem(E,\Xmax,\theta)$ to determine the electromagnetic contribution
to $S(1000)$. The remainder of the signal is attributed to muons, and
we can determine a muon normalization for each event:
\begin{equation}
\Nmu = \frac{S(1000) - \Sem(E, \Xmax, \theta)}{\Smu(\theta, \Xmax)},
\end{equation}
where $\Smu$ is again the reference muon signal (proton-QGSJetII at 10~EeV).

%%%%%%%%%%%%%%%%%%%%%%%%%%%%%%
\begin{figure}[t]
\includegraphics[width=\psize]{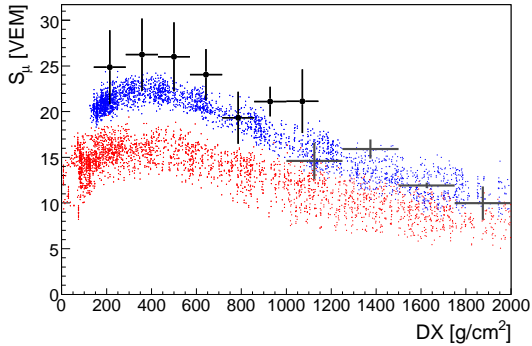}
\caption{\small{Reconstructed muon tank signal contribution
as a function of shower evolution (distance to ground $DX$)
for hybrid events with $\theta < 60^\circ$ (black points) and
$60\Deg < \theta < 68\Deg$ (grey points), using an energy scale of
$E= 1.26 E_{\rm FD}$. The muon signal from simulated
proton (red) and iron (blue) showers (using QGSJetII) is also shown.
\label{fig:hybrid-muons}}}
\end{figure}

%%%%%%%%%%%%%%%%%%%%%%%%%%%%%%

We select high-quality hybrid events for which the
shower maximum is in the field of view of a telescope, $\theta <
60^\circ$, and the Mie scattering length is measured. Furthermore, the 
Cherenkov light fraction is limited to less than 50\%, and we apply all 
cuts used in the measurement of the elongation
rate \cite{Auger-ICRC07-Unger} which have been carefully chosen in order to ensure an
unbiased $\Xmax$ distribution of the air showers. 
The surface detector event has to satisfy the T5
selection cuts which are also applied in \cite{PRL}.

%%%%%%%%%%%%%%%%%%%%%%%%%%%%%%
\begin{figure}[thb!]
\centerline{
\includegraphics[width=\psize]{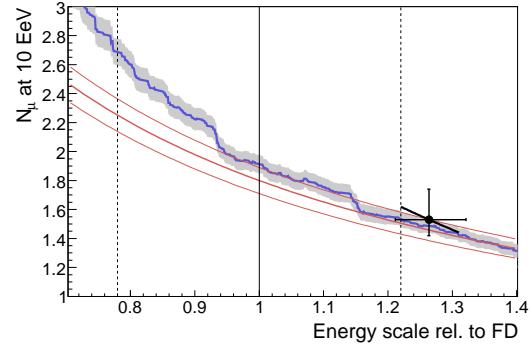}
}
\caption{\small{Comparison of the results on 
$\Nmu$ at 10\,EeV from different methods. The black dot shows the
constant intensity method result with statistical (diagonal line) and
systematic errors (vertical/horizontal lines). The blue
line with shaded band, and red lines indicate the result from vertical 
and inclined hybrid events, respectively, both with statistical errors. 
The vertical dashed lines indicate the systematic
error on the fluorescence energy scale \cite{PRL}.
\label{fig:comparison}}}
\vspace*{-0.5cm}
\end{figure}
%%%%%%%%%%%%%%%%%%%%%%%%%%%%%%

The average muon normalization in hybrid events at 10\,EeV is found to be
\begin{eqnarray}
\left. \Nmu\right|_{E= 1.26 E_{\rm FD}} &=& 1.49 \pm 0.05, \;\mbox{and}
\nonumber\\
\left. \Nmu\right|_{E= E_{\rm FD}} &=& 1.84 \pm 0.07,
\end{eqnarray}
in excellent agreement with the constant intensity method results, when
compared at the same energy scale. 

A similar study has been performed for inclined hybrid events
($60^\circ < \theta < 68^\circ$). While the statistics are 
limited, no subtraction of $\Sem$ is necessary for these events,
as the electromagnetic part has been attenuated in the large depth of 
atmosphere traversed. Thus, inclined hybrid events yield a clean
measurement of $\Nmu$, and good
agreement between muon numbers of the inclined and the vertical data
sets is found.

Fig.~\ref{fig:hybrid-muons} shows the muon signal derived from
the hybrid events as a function of shower evolution ($DX$). While the
muon signal is clearly higher than that predicted in the simulations,
the behavior as a function of $DX$ follows the prediction very well,
which serves as a consistency test of our method.

In Fig.~\ref{fig:comparison} we compare the results of the different
methods applied for inferring the muon density at 1000\,m from the
shower core. The relative number of muons is shown as function of the
adopted energy scale with respect to the Auger fluorescence detector
energy reconstruction. Only the constant intensity method yields an
independent measurement of the energy scale.
Good agreement between the presented methods is found, when compared
at $E \sim 1.26\:E_{FD}$.

%%%%%%%%%%%%%%%%%%%%%%%%%%%%%%%%%%%%%%%%%%%%%%%%%%%%%%%%%%%%%%%%%%%%%%%%%%%
\section{Discussion}
\label{sec:disc}

Assuming air shower universality, which was tested with simulations
of various primaries and hadronic interaction models,
we have determined the muon density at 10~EeV and the
energy scale with which the data of the Pierre Auger Observatory can be
described self-consistently. The number of muons measured in data is
about 1.5 (1.7) times larger than that predicted by QGSJET II (Sibyll) for 
proton showers, with statistical and systematic errors of the order of 0.1. 
This value of $\Nmu$ is even slightly larger than that predicted for
iron showers, by a factor of 1.1 and 1.2 for QGSJetII and Sibyll, respectively.
A similar excess has been found at energies from 3~EeV to 50~EeV.

Consistent results were obtained with several analysis
methods on independent data sets: the constant intensity method (\refsec{CIC})
uses only the surface detector data and is independent of the fluorescence
energy scale. Vertical hybrid events (\refsec{hybrid}) allow for a direct
and unbiased measurement of $\Nmu$ for each event, albeit depending on the
flourescence energy. The signal in inclined hybrid events ($\theta > 60\Deg$)
is purely muonic, and the resulting $\Nmu$ independent of the subtraction
of the electromagnetic signal. The fact that all three methods agree within
errors for an energy scale of $1.26 E_{FD}$ indicates that increasing the
number of muons by 50\% (relative to proton-QGSJetII) yields a consistent 
description of the Auger data.

If interpreted in terms of cosmic ray composition, the derived muon
density would correspond to nuclei heavier than
iron, which is clearly at variance with the measured $X_{\rm max}$
values at this energy. The discrepancy between air shower data and simulations
reported here is qualitatively similar to the inconsistencies found in
composition analyses of previous detectors
\cite{Abu-Zayyad:1999xa,Dova:2004nq,Watson:2004rg}.
This points towards a deficiency of hadronic models in predicting the number
of muons produced in air showers at large distances from the shower core.
However, it is important to note that the method presented here relies on a 
correct description of the electromagnetic shower component in the simulations.

%Recently, the EPOS model...{\bf comments on EPOS ?}
\begin{small}

\end{small}

% \begin{thebibliography}{99}
% \end{thebibliography}

\end{document}